\documentclass[11pt,twoside]{article}

\usepackage{asp2006}
\usepackage{epsf}
\usepackage{lscape}

\begin{document}
\title{Refining the Radius--Luminosity Relationship for Active Galactic Nuclei}
\author{Misty~C.~Bentz, Kelly~D.~Denney, Bradley~M.~Peterson, and Richard~W.~Pogge}
\affil{The Ohio State University Department of Astronomy\\
       140 W. 18th Ave., Columbus, OH 43210 USA} 

\begin{abstract}

We have measured the host-galaxy starlight contribution to four
lower-luminosity AGNs (NGC~3516, NGC~4593, IC~4329A, and NGC~7469).  We
include these objects with new broad line region measurements for
NGC~4151 and NGC~4593 to present a revised version of the
radius--luminosity relationship for AGNs.

\end{abstract}

 \section{Introduction}

The radius--luminosity ($R - L$) relationship is an extremely useful
product that results from years of reverberation-mapping (variability)
campaigns to study AGNs.  It is the basis for estimates of black hole
masses in objects where direct measurements are either not feasible or
not practical.

Recently, \citet{bentz06a} presented a revised version of the $R - L$
relationship based on starlight-corrected luminosities for
reverberation-mapped AGNs based on high-resolution {\it Hubble Space
Telescope (HST)} imaging of the galaxy centers.  The power law slope
determined for this revised relationship was $\alpha = 0.52$, shallower
than previously determined and consistent with naive photoionization
arguments.  This initial sample of starlight-corrected objects was 14,
and several of the lower-luminosity objects that did not have imaging
available were excluded from the fit of the $R - L$ relationship.  In
this work, we present an additional four objects (NGC~3516, NGC~4593,
IC~4329A, and NGC~7469) with high-resolution {\it HST} imaging and
measurements of the host-galaxy starlight contribution to their
luminosity measurements.  We also include new broad line region (BLR)
radius measurements from the recent reverberation-mapping campaigns of
NGC~4151 \citep{bentz06b} and NGC~4593 \citep{denney06}.  We reanalyze
the fit to the $R - L$ relationship and present the resulting formula
for estimating black hole masses based on the new $R - L$ calibration.

 \section{Observations and Host-Galaxy Flux Measurements}

{\it HST} imaging of NGC~3516, NGC~4593, IC~4329A, and NGC~7469 was
obtained throughout Cycle 14 (2005-2006).  The observations, reduction
methods, and galaxy fitting methods were similar to those described by
\citet{bentz06a}, but for completeness, we include a short description
here.

Each object was observed with the Advanced Camera for Surveys (ACS) High
Resolution Channel through the F550M filter (medium-band $V$).  A set of
graduated exposure times (120~s, 300~s, and 600~s) was employed to both
acquire unsaturated images of the nucleus but also to achieve a
reasonable signal-to-noise ratio in the outlying galaxy.  Saturated
pixels in the longer exposures were replaced with unsaturated pixels
from shorter exposures that had been scaled by the exposure time
difference.  The corrected images were stacked and cleaned of cosmic
rays and then tranformed to account for the distortions of the ACS
camera.

\begin{table}[!ht]
\caption{Observed Galaxy Fluxes and AGN Corrected Luminosities}
\smallskip
\begin{center}
{
\small
\begin{tabular}{cccccc}
\tableline
\noalign{\smallskip}
Object & Aperture & PA     & $f_{\lambda, \rm gal}$ (5100\AA) & $\lambda L_{\lambda, \rm AGN}$ (5100\AA) & Refs.\\
       & {\footnotesize (\arcsec\ $\times$ \arcsec)}  & {\footnotesize (\deg)} & {\footnotesize ($10^{15}$ ergs s$^{-1}$ cm$^{-2}$ \AA$^{-1}$)} & {\footnotesize ($10^{44}$ ergs s$^{-1}$)} & \\
\noalign{\smallskip}
\tableline
\noalign{\smallskip}
NGC 3516 & $1.5 \times 2$   & 25   & $4.64 \pm 0.86$   & $0.032 \pm 0.026$ & 1,2 \\
NGC 4593 & $5 \times 12.75$ & 90   & $11.54 \pm 2.14$  & $0.044 \pm 0.022$ & 3   \\
IC 4329A & $5 \times 10$    & 90   & $4.90 \pm 0.91$   & $0.032 \pm 0.038$ & 4   \\
NGC 7649 & $10 \times 16.8$ & 26.7 & $15.98 \pm 2.96$  & $0.135 \pm 0.129$ & 5   \\
\noalign{\smallskip}
\tableline
\end{tabular}
}
\end{center}
{\small
References: 1. \citet{onken03}, 2. \citet{wanders93}, 3. \citet{denney06},
	    4. \citet{winge96}, 5. \citet{collier98}.\\
}
\end{table}

The final images were fit with two dimensional galaxy models using
Galfit \citep{peng02}.  A simultaneous fit of the disk, bulge, and
central PSF was determined for each object.  Once a satisfactory fit was
found, the central PSF was subtracted, leaving a nucleus-free image of
each galaxy.  The original ground-based monitoring aperture was overlaid
on each host galaxy image and the starlight within the aperture was
summed to give the host-galaxy contribution to the flux measured for
each AGN.  Table~1 lists the aperture geometries and orientations for
each of the four objects, along with the measured host galaxy flux and
the resulting starlight-free luminosity of the AGN.

 \section{\boldmath Recalibrating the $R - L$ Relationship}

We combine the corrections above with those of \citet{bentz06a} and also
replace previous reverberation-mapping results for the radii of the BLRs
in NGC~4151 and NGC~4593 with the new results determined by
\citet{bentz06b} and \citet{denney06}.  Multiple measurements for the
same object were combined into a weighted mean.  We then used the
orthoganal least-squares analysis package GaussFit \citep{mcarthur94} to
determine the $R - L$ relationship for H$\beta$,

\begin{equation}
\log R_{\rm BLR} = -22.198 + 0.539 \log \lambda L_{\lambda} {\rm (5100 \AA)}
\end{equation}

\noindent where $R_{\rm BLR}$ is the average radius of the H$\beta$ BLR.  
This recalibration of the $R - L$ relationship is not significantly
different from that determined by \citet{bentz06a}, even with the
inclusion of additional data points.  Figure~1 shows the calibration
determined above contrasted with the calibration determined by
\citet{kaspi05}, which does not include host-galaxy starlight
corrections.

\begin{figure}[!ht]
\plotone{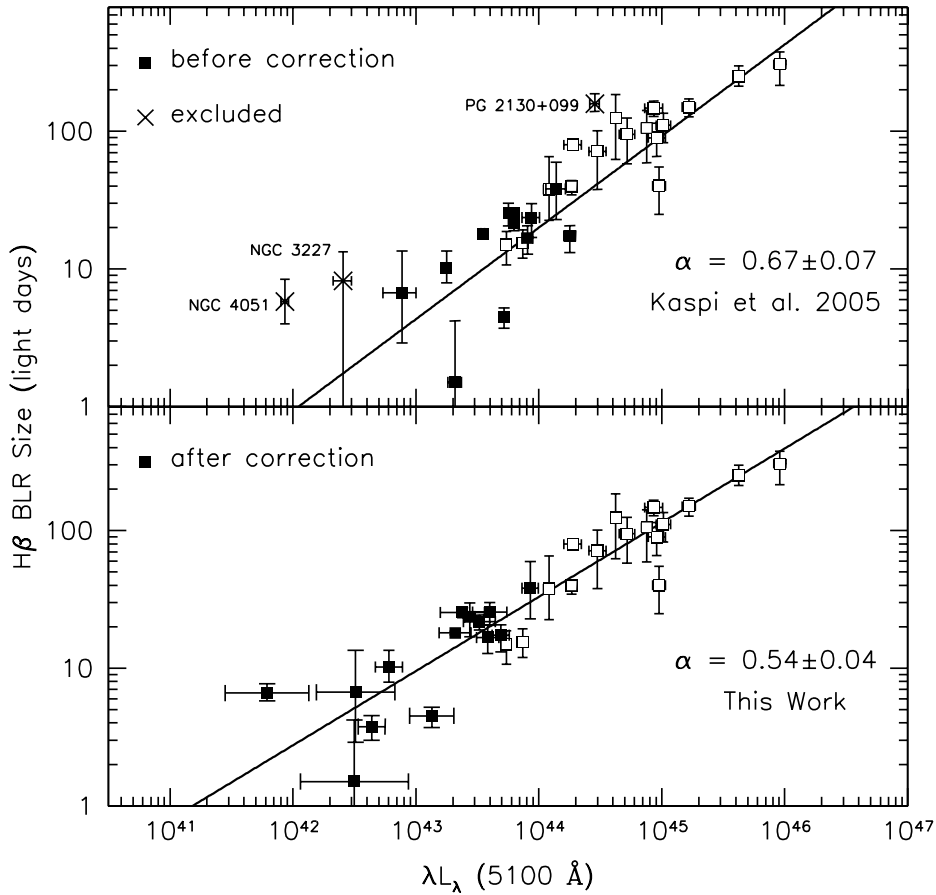}
\caption{The top panel shows the recalibration of the $R - L$ relationship 
         from \citet{kaspi05} using the reanalyzed reverberation
         results of \citet{peterson04}.}
\end{figure}

Removing the host-galaxy starlight component significantly reduces the
scatter in the relationship, but also flattens the slope of the
relationship considerably.  This has the overall effect of biasing
samples that use the $R - L$ relationship to estimate black hole masses.
The largest effect is for low-luminosity objects where black hole masses
could be overestimated by a factor of $> 3$.  The host-galaxy starlight
is typically removed from flux measurements of low-luminosity objects
before the black hole mass is estimated.  However, it is also crucial to
use a $R - L$ relationship that has been determined after removal of the
host-galaxy starlight from the population of objects providing the
calibration.

To estimate black hole masses, the luminosity of the object is used in
combination with the $R - L$ relationship to estimate the radius of the
BLR.  The following equation gives the combination of the above
determination for the $R - L$ relationship and the formula for
calculating black hole masses:

\begin{equation}
\log M_{\rm BH} = 0.808 + 0.539 \log L_{44} + 2 \log V + \log f.
\end{equation}

\noindent Here, $L_{44}$ is the luminosity of the object in units of
$10^{44}$~ergs~s$^{-1}$, $V$ is the velocity width of the H$\beta$
emission line in km~s$^{-1}$, and $f$ is a geometric factor.
\citet{onken04} find that $\langle f \rangle = 5.5$ for the
reverberation-mapped objects, where $V$ is measured as the second moment
of the variable (RMS) line profile, $\sigma_{\rm line}$.  While this
measurement of the line width is not possible for single epoch spectra,
$\sigma_{\rm line}$ may still be calculated from the mean line profile,
and again $f = 5.5$ should be used.  It is important when estimating
black hole masses using the above equation to correct any FWHM
measurements of the line profile by using the $f$ values determined by
\citet{collin06} in their eq. 7.

 \section{Conclusions}

Removing the host-galaxy contribution to luminosity measurements of
re\-ver\-ber\-ation-mapped AGNs substantially decreases the slope of the
$R - L$ relationship.  Failing to take the host galaxy starlight into
account will serve to bias estimates of black hole masses that are based
on the $R - L$ relationship, especially on the low luminosity end where
the masses can be overestimated by a factor of $> 3$.  The $R - L$
relationship has now been shown to hold across five orders of magnitude,
with fairly low scatter.  As such, it is a powerful diagnostic tool in
that outliers from the relationship can be expected to have physical
differences that cause them to be separated from the general population
of AGNs.

\acknowledgements 
This work is based on observations with the NASA/ ESA {\it Hubble Space
Telescope}.  We are grateful for support of this work through grants
{\it HST} GO-9851 and GO-10516 from the Space Telescope Science
Institute, which is operated by AURA, under NASA contract NSA5-26555,
and by the NSF through grants AST-0205964 and AST-0604066 to The Ohio
State University.  M.C.B. is supported by a National Science Foundation
Graduate Fellowship.


\begin{thebibliography}{}

\bibitem[{{Bentz} {et~al.}(2006a){Bentz}, {Peterson}, {Pogge},
  {Vestergaard}, \& {Onken}}]{bentz06a}
{Bentz}, M.~C., {et~al.} 2006a, \apj, 644, 133

\bibitem[{{Bentz} {et~al.}(2006b)}]{bentz06b}
{Bentz}, M.~C., {et~al.} 2006b, \apj, 651, 775

\bibitem[{{Collier} {et~al.}(1998)}]{collier98}
{Collier}, S.~J., {et~al.} 1998, \apj, 500, 162

\bibitem[{{Collin} {et~al.}(2006){Collin}, {Kawaguchi}, {Peterson}, \&
  {Vestergaard}}]{collin06}
{Collin}, S., {et~al.} 2006, \aap, 456, 75

\bibitem[{{Denney} {et~al.}(2006)}]{denney06}
{Denney}, K., {et~al.} 2006, \apj, 653, 152

\bibitem[{{Kaspi} {et~al.}(2005)}]{kaspi05}
{Kaspi}, S., {et~al.} 2005, \apj, 629, 61

\bibitem[{{McArthur} {et~al.}(1994)}]{mcarthur94}
McArthur, B., Jefferys, W., \& McCartney, J.  1994, \baas, 26, 900

\bibitem[{{Onken} {et~al.}(2003)}]{onken03}
{Onken}, C.~A., {et~al.} 2003, \apj, 585, 121

\bibitem[{{Onken} {et~al.}(2004)}]{onken04}
{Onken}, C.~A., {et~al.} 2004, \apj, 615, 645

\bibitem[{{Peng} {et~al.}(2002)}]{peng02}
{Peng}, C.~Y., Ho, L.~C., Impey, C.~D., \& Rix, H. 2002, \aj, 124, 266

\bibitem[{{Peterson} {et~al.}(2004)}]{peterson04}
{Peterson}, B.~M., {et~al.} 2004, \apj, 613, 682

\bibitem[{{Wanders} {et~al.}(1993)}]{wanders93}
{Wanders}, I., {et~al.} 1993, \aap, 269, 39

\bibitem[{{Winge} {et~al.}(1996)}]{winge96}
{Winge}, C., {et~al.} 1996, \apj, 469, 648

\end{thebibliography}
\end{document}